\documentclass[prb, twocolumn, amssymb, amsmath,aps,superscriptaddress,showpacs]{revtex4}
\usepackage{natbib}
\usepackage{colordvi}
\usepackage[dvips]{color}
\usepackage{graphicx}
\usepackage{epstopdf}
\usepackage{bm}
\usepackage{amsmath, amsfonts}

\begin{document}

\title{Electronic and magnetic properties in strongly correlated heterostructures}
\author{Suguru Ueda}
\affiliation{Department of Physics, Kyoto University, Kyoto 606-8502, Japan}
\author{Norio Kawakami}
\affiliation{Department of Physics, Kyoto University, Kyoto 606-8502, Japan}
\author{Manfred Sigrist}
\affiliation{Theoretische Physik, ETH Z\"urich, CH-8093 Z\"urich}

\date{\today}                                           

\begin{abstract}
We present a theoretical study of a model heterostructure for a Mott-insulator sandwiched between two band insulators, such as SrTiO$_3$/LaTiO$_3$. Particular emphasis is given on the interplay between magnetism and inhomogeneous charge distributions. Our mean-field analysis of the generalized Hubbard model displays numerous ordered phases in the ground-state phase diagram. In particular, we find a canted antiferromagnetic state near the interface when antiferromagnetic-ordering exists inside the Mott insulator. A checkerboard charge-ordering proposed previously is also stabilized for large long-range Coulomb interactions. Regarding its origin we also point out the importance of interlayer spin-mediated interactions. It is further shown that such a strong spin-charge coupling gives rise to pronounced magnetic/charge order phase transitions in external magnetic fields: a first-order metamagnetic transition and a reentrant charge-order transition with checkerboard pattern. The mechanisms stabilizing these intriguing phases are explored through a detailed analysis of the physical quantities with special focus on the spin-charge interplay.
\end{abstract}
\maketitle

\section{Introduction}

Correlated heterostructures composed of transition-metal oxides have attracted much interest both experimentally and theoretically as a new arena for studies of strongly correlated electron systems.
Especially, recent progress in crystal growth techniques makes it possible to fabricate and control several classes of oxide heterojunctions. Since Ohotomo {\it et al.} found interface-specific conducting states between two different insulating perovskite transition-metal oxides~\cite{ohtomo:2002, nature_427}, one of the main interests has been directed toward the electronic properties near the hetero-interface, drastically varying with respect to the bulk components. One of the best-known examples is the LaAlO$_3$/SrTiO$_3$ heterostructure. Although both LaAlO$_3$ and SrTiO$_3$ are conventional band-insulators, recent experiments have revealed two-dimensional superconductivity~\cite{sience.317.1196}, ferromagnetic correlation~\cite{nmat493} and even the coexistences of the two states at $n$-type polar interfaces~\cite{1105.0235, PhysRevLett.107.056802}. These phases exhibit a distinct dependence on film-thickness and growth-condition~\cite{nmat493,kalisky2012critical} and an intriguing electric-field response~\cite{PhysRevLett.104.126802, PhysRevLett.104.126803}. Furthermore, high carrier-mobility at the interface~\cite{ohtomo:2002,takizawa:pl,PhysRevB.73.195403,PhysRevLett.99.266801} and two-dimensional superconductivity~\cite{Biscaras:2010:nture.com, biscaras2011two} have been reported for LaTiO$_3$ (LTO)/SrTiO$_3$ (STO) heterostructures. 

Motivated by these experimental findings, many theoretical approaches have been proposed to describe the nature of quasi two-dimensional electron systems localized around the interface~\cite{PhysRevB.74.075106, PhysRevB.74.195427, Okamoto:2007, PhysRevB.77.115350,PhysRevB.79.045130,PhysRevB.76.075339}. Particularly the LTO/STO interface has been extensively investigated due to the chemical similarity and small mismatch of lattice constants between perovskite compounds LTO and STO~\cite{okamoto:2004:hf}. Okamoto and Millis analyzed a generalized Hubbard model in the framework of Hartree-Fock approximation~\cite{okamoto:nature,okamoto:2004:hf} and dynamical mean field theory (DMFT)~\cite{okamoto:2004:dmft, okamoto:2005}. The origin of metallic behavior at the LTO/STO interface has been clarified, and, moreover, possible ferromagnetic and antiferromagnetic phases have been investigated, also in relation to orbital ordering~\cite{okamoto:2004:hf} and their temperature dependence~\cite{okamoto:2005}. First principles calculations have been also performed to explore the role of the lattice reconstruction, and demonstrated several spin and charge ordered phases~\cite{okamoto:dft,pentcheva:2007} as well as the effects of lattice relaxation~\cite{PhysRevB.73.195403}, which modify the interface electronic structure. Moreover, R\"uegg {\it et al} examined the electronic and thermoelectric transport properties, particularly focusing on the characters of quasi-particles in the inhomogeneous layered system~\cite{ruegg:2008}.

In this paper, we present a theoretical study of electronic and magnetic properties in correlated heterostructures of the Mott-insulator (MI) embedded between two band insulators (BI), such as STO/LTO/STO structures. A sketch of our model system is shown in Fig.~\ref{fig:model}. According to the previous optical and photoemission studies, LTO/STO heterojunction develops a quasi-two-dimensional electron gas extended only a few unit-cells across the interface. As mentioned in Ref.~\onlinecite{PhysRevB.82.201407}, however, some experimental setups may allow us to tune the dimensionality of the interface electron system. Therefore, reflecting the broadness of the conducting electronic state, the possible $d$ band occupation would vary abruptly or gradually from $n=1$ ($d^1$) at LTO to $n=0$ ($d^0$) at STO region beyond the interface if the thickness of the heterostructure along the stacking direction is large enough to reproduce bulk properties in each region. Such sharpness of the spatially changing electron density might even modify the electronic structure only near the interface. 

Here we shed light on the nature of the electronic state around the hetero-interface with particular emphasis on the interplay of charge distribution and magnetic properties. As mentioned above, we consider a BI/MI/BI heterostructure, taking the example of a LTO/STO heterostructure. For simplicity, we assume perfect matching of the two lattices and, thus, only focus on the electronic reconstruction based on the mobile electrons in the Ti-derived band. The corresponding model heterostructure is described by the generalized Hubbard model defined in Ref.~\onlinecite{okamoto:2004:hf}. Here our focus is put on the competition of spin and charge degrees of freedom while neglecting the orbital degeneracy of Ti-3d band. We treat on-site interaction with the Hartree-Fock approximation, and calculate the ground state phase diagram as a function of the local and long-range Coulomb interaction. We obtain a variety of the interface-specific magnetic and charge-ordered phases. We find that the intimate spin-charge coupling in magnetic fields causes a first-order metamagnetic transition as well as a reentrant charge-ordering transition. 

This paper is organized as follows. In Sec.~\ref{sec:model}, we introduce the model and method employed in this study. In Sec.~\ref{sec:phase diagram} we discuss the $T=0$ phase diagram obtained as a function of local and long-ranged Coulomb interactions, and provide detailed studies of some ordered phases specific to the interface. Sec.~\ref{effects of magnetic fields} presents the effects of external magnetic fields, and clarifies the origin of field-induced spin and charge transitions. Finally, a brief summary is given in Sec.~\ref {sec:conclusion}.

\section{Model} \label{sec:model}

\begin{figure}
\centering
\includegraphics[width=0.8\linewidth]{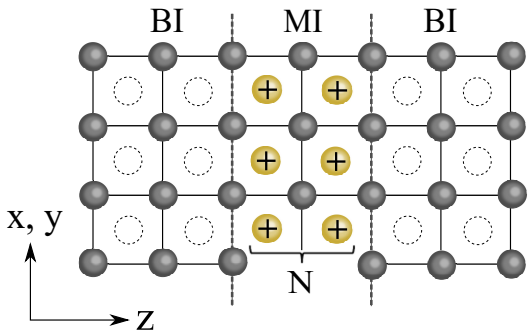}
\caption{Schematic view of the BI/MI/BI-type [001] heterostructure defined by $N$ layers of positively charged cations. We choose the $z$-axis along $[001]$ direction. Black sites at $\vec{R}_i =a(l_i, m_i, n_i)$ correspond to the $B$ sites of perovskite $ABO_3$ and $A'BO_3$ lattices, where the relevant conduction electrons reside. The $A$ and $A'$ ions located at $\vec{R}_i^{ion} =a(l_i+1/2, m_i+1/2, n_i+1/2)$ are replaced by $+1$ and neutral counter-ions, respectively.}
\label{fig:model}
\end{figure}

We study a BI/MI/BI-type heterostructure based on the [001] stacking of $d^1$ Mott insulator $ABO_3$ and $d^0$ band insulator $A'BO_3$ with cubic perovskite lattice structures; hereafter we choose the $z$ axis along the [001] direction. Taking an example of LTO/STO multilayer structure, $A$ ($A'$) ions represent a valence of +2 (+3) charged cation, and mobile conduction electrons reside on the $B$ ions which are common constituents for $ABO_3$ and $A'BO_3$. A schematic view of this system is illustrated in Fig.~\ref{fig:model}. We assume the chemical similarity and perfect lattice match between two cubic perovskite components. Therefore, simulating the different valences between $A$ and $A'$ ions, we define the model heterostructure via $N$ counter-ion layers with a positive charge $+e$ sitting in the Mott-insulating material; $A^{3+}$ and $A'^{2+}$ cations are here substituted with $+1$ and neutral point charges, respectively~\cite{okamoto:2004:hf}. Hence, in addition to the repulsive electron-electron interaction, electrons suffer from the electrostatic potential of the positively charged ions placed between electrically active $B$ sites. The charge neutral condition requires that the total number of these electrons equals to that of $+1$-charged $A$ site ions. In the present work, the corresponding microscopic model is simplified with disregarding orbital degeneracy of $B$ ions, which are sometimes associated with interesting phenomena in transition-metal oxides. Thus, this model heterostructure captures the essential aspects of the competition between spin and charge degrees of freedom in a spatially non-uniform environment.

To describe electron dynamics in the background of the non-uniform attractive potential, we introduce the generalized Hubbard model given by,
\begin{equation}
H = H_{band} +H_{ee} +H_{ei}. \label{ham}
\end{equation}
%
Here, the kinetic term $H_{band}$ is presented by the single band tight-binding model,  
%
\begin{equation}
H_{band} = -t \sum_{ <i,j>, \sigma } ( c_{i \sigma} ^{\dagger} c_{j \sigma} +h.c. ), \label{kin}\\
\end{equation}
%
where $c_{i \sigma}$ is an annihilation operator of an electron with spin $\sigma = \uparrow ,\downarrow $ at the site labeled by index $i$ as $\vec{R}_i =a(l_i, m_i, n_i)$ with lattice constant $a$. Electron hopping with the transfer integral $t$ is limited between the nearest-neighbor sites on the square lattice. $H_{ee}$ and $H_{ei}$ correspond to electron-electron and electron-ion Coulomb interaction given by,
%
\begin{eqnarray}
H_{ee}  &=&  U\sum_{i} \hat n_{i \uparrow } \hat n_{i \downarrow } + \frac{1}{2}\sum_{i \not= j, \sigma, \sigma '}  \frac{ e^2 \hat n_{i \sigma} \hat n_{j \sigma '} } { \epsilon | \vec{R}_i - \vec{R}_j | }, \label{ee}\\
H_{ei}  &=& - \sum_{i j, \sigma }  \frac{ e^2 \hat n_{i \sigma} } { \epsilon | \vec{R}_i - \vec{R}_j^{ion} | }, \label{ei}
\end{eqnarray}
%
where $n_{i \sigma} = c^{\dagger}_{i \sigma} c_{i \sigma}$ is the spin dependent occupation number, and $\vec{R}_i^{ion} =a(l_i+1/2, m_i+1/2, n_i+1/2)$ specifies a position of the cation. Both the electron-electron repulsion and the electron-ion attraction are introduced as a long-ranged Coulomb interaction whose characteristic energy scale is given by the coefficient $E_c = e^2/a\epsilon$ with the dielectric constant $\epsilon$ of the host lattice. In Eq. (\ref{ee}), the strong on-site repulsion between electrons is considered via an ordinary Hubbard interaction term with magnitude $U$. Note that the strength of the on-site interaction has the same value on all $B$ sites throughout the heterostructure, i.\ e.\ the spatial inhomogeneity of this model is characterized by the $H_{ei}$ term. Therefore, it is naturally assumed that the electron density profile is mainly determined by these long-range interactions. On the one hand, the broad charge distribution, extended into the band-insulating material, is favored in order to suppress repulsive electron-electron interaction $H_{ee}$, but on the other hand, the attractive electrostatic potential $H_{ei}$  confines the conduction electrons in the Mott-insulating region. 

We also take the effects of external magnetic fields into account to give a systematical study of the magnetism in the inhomogeneous system. In the current study, the magnetic fields are treated through the Zeeman term. 
%
\begin{equation}
H_{Z} = - \frac{1}{2}g \mu _B H \sum_{i} m_{iz}, \label{zeeman}
\end{equation}
%
where $g$ and $\mu_B$ indicate the electron g-factor and the Bohr magneton, respectively, and $m_{iz} = n_{i\uparrow} -n_{i\downarrow }$ is a magnetization along the applied magnetic fields. Hereafter, the applied magnetic fields are measured in units of $\frac{1}{2} g \mu _B$.  

The ground state phases are governed essentially by three parameters: thickness of positive ion layers $N$, on-site Hubbard $U$ and the coupling constant of long-range Coulomb interaction $E_c$. As mentioned in Ref.~\onlinecite{okamoto:2004:hf}, for the example of the experimentally fabricated LTO/STO heterostructure substantial uncertainties exist in estimating the realistic values of $U$ and $E_c$ except for an experimentally tunable parameter $N$. Particularly the dielectric constant $\epsilon = e^2/aE_c$ in bulk STO samples strongly depends on temperature~\cite{PhysRevLett.26.851,PhysRev.155.796} and electric fields~\cite{PhysRevLett.102.216804}, and can even change its value by more than one order of magnitude as a function of these quantities. Therefore, in addition to $N$, discussions on the role of $U$ and $E_c$ may be important to gain physical insights into the electronic and magnetic properties of realistic heterostructures, particularly into the interplay of spin and charge degrees of freedom. However, a systematic and exact computational study of the model Hamiltonian [Eq.~(\ref{ham})] over a wide range of parameters is practically excluded. Hence, in the current study, the on-site repulsion and long-range Coulomb interaction are treated by employing unrestricted Hartree-Fock and Hartree approximation, respectively. Here, for the derivation of the effective single-particle Hamiltonian, the on-site Hubbard term $U\sum_{i}n_{i\uparrow }n_{i\downarrow }$ in Eq. (\ref{ee}) decoupled as follows,
%
\begin{eqnarray}
n_{i\uparrow }n_{i\downarrow } &\stackrel{HF}{\longrightarrow}& \hat n_{i \uparrow } \langle \hat n_{i \downarrow  } \rangle 
            +\langle \hat n_{i \uparrow } \rangle \hat n_{i \downarrow }   
			-\langle \hat n_{i \uparrow } \rangle \langle \hat n_{i \downarrow } \rangle \nonumber\\
		& &{}- \langle c_{i \uparrow }^{\dagger} c_{i \downarrow } \rangle c_{i \downarrow  }^{\dagger} c_{i \uparrow }    
		    -c_{i \uparrow }^{\dagger} c_{i \downarrow }\langle c_{i \downarrow  }^{\dagger} c_{i \uparrow } \rangle  \nonumber\\
		& &{}+ \langle c_{i \uparrow }^{\dagger} c_{i \downarrow } \rangle  \langle c_{i \downarrow  }^{\dagger} c_{i \uparrow } \rangle
\end{eqnarray}
%
where the site-dependent expectation values $\langle \dots \rangle$ are calculated self-consistently. However, not only the energetically most favorable state, several self-consistent solutions are typically allowed in mean-field treatments. We thus consider a variety of ordered states allowed up to two sublattices and select the one with the lowest energy after many iteration steps of $O(10^3)$. As compared to other many-body methods such as dynamical mean field theory, the Hartree-Fock treatment treats electron correlation effects only with limitations. We, however, believe that the approximation captures the qualitatively correct tendency towards certain electron correlations, while overestimating
the realization of actual order. 

In the following, we restrict our investigation to the physically relevant parameter regime $U>E_c$ at zero temperature. We consider a supercell geometry of about $25$-layers structure including $N$ cation planes, and employ open boundary conditions for solving the mean-field Hamiltonian. Since electrons are well confined within this multi-layer structure for all the parameter values considered in our study, the boundary condition would not change the quantitative aspects of our present calculation.

\section{Phase diagram} \label{sec:phase diagram}

In order to gain insights into the electronic and magnetic properties, we first discuss the ground-state phase diagram. In this section, we fix the external magnetic field to zero and investigate possible phases in the $U$-$E_c$ plane. As pointed out by previous studies~\cite{okamoto:2004:hf,lin:2005}, the profile of the charge distribution is sensitive to the energy scale of long-range Coulomb interaction $E_c$. At small $E_c$, electrons are extended to the band-insulating region due to the weak attractive potential of cation layers, and the electron density profile gradually varies from $n\simeq1$ (MI) to $n\simeq 0$ (BI) across the interface. In contrast, electrons are strongly confined along the $z$ direction for large $E_c$, and a more abrupt change of charge distribution is observed in the vicinity of the interface. To characterize this ``sharpness" in the charge distribution around the interface, we introduce a Thomas-Fermi screening length which roughly corresponds to $\lambda_{TF}\sim\sqrt{t/E_c}a$~\cite{ruegg:2008}. With this value, we can distinguish between ``broad" and ``abrupt" variations of the charge distribution via $\lambda_{TF}<a$ and $\lambda_{TF}>a$, for which the positive electrostatic potential confines mobile electrons between the insulating regions. We expect that the electronic and magnetic phases near the interface are sensitive to the electron density profile, and change their character around $\lambda_{TF}\simeq a$. Similar insights have been also given in the preceding studies using double exchange ~\cite{lin:2005} and Holstein-Hubbard model~\cite{nanda:2011}.

We investigate possible in-plane magnetic and charge ordered states with a two-sublattice structure. DMFT calculations performed by Okamoto and Millis~\cite{okamoto:2005} showed that a ferromagnetic ordering is favored in the vicinity of the interface in the large $U$ regime. Therefore, besides the conventional paramagnetic (PM), ferromagnetic (F) and antiferromagnetic (AF) states, we allow for a spatial modulation of the magnetic moment. Along the $z$-axis, a possible magnetic state can vary smoothly from the AF state in the Mott-insulating region to the FM state around the interface. We search for the favorable direction of the FM moment relative to that of the AF moment. In addition, although a simple cubic lattice with band filling $\simeq 1$ prefers the G-type AF (GAF) ordering showing $(\pi ,\pi ,\pi )$ spin configuration particularly in the $N\rightarrow \infty$ limit, the strong spatial inhomogeneity in the charge density profile may induce a uniaxial distortion for the possible spin alignment in the $z$-direction. Hence, as trial magnetic states in the Mott-insulating region, we further take into account the A-type AF (AAF) and C-type AF (CAF) structure having the $(0,0,\pi )$ and $(\pi ,\pi , 0)$ spin orderings, respectively.

%
\begin{figure}
\centering
\includegraphics[width=1.0\linewidth]{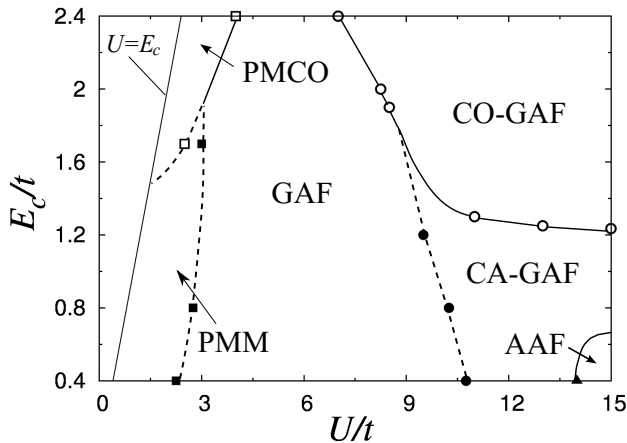}
\caption{Ground-state phase diagram as a function of the on-site repulsion $U$ and long-range Coulomb interaction $E_c$. The number of the counter-ion layers is  $N=4$. Our focus is put on the physically relevant regime, $U>E_c$. Six phases are shown: paramagnetic  metallic state (PMM), paramagnetic charge ordered state (PMCO),  antiferromagnetic state with $(\pi, \pi, \pi)$ spin configuration  (GAF), $(0, 0, \pi)$-type antiferromagnetic state (AAF), GAF state with canted antiferromagnetic state at the interface (CA-GAF) and GAF state with checkerboard charge-ordering at the interface (CO-GAF). The solid  and broken lines indicate the first- and second-order transitions, respectively. } 
\label{fig:n4_UEc_Phase}
\end{figure}
%

Fig.~\ref{fig:n4_UEc_Phase} presents the calculated phase diagram for an $N=4$ layer structure with counter-ions at $z=\pm 1.5$ and $z=\pm0.5$. To obtain the phase diagram, we compare the energies of several different ordered phases and select the one with the lowest energy. Six phases are found within the current mean-field analysis: PM metallic state (PMM), PM charge ordered (CO) state with checkerboard pattern (PMCO),  GAF state, AAF state, canted AF state around the interface accompanied with GAF ordering in Mott-insulating region (CA-GAF), and ferromagnetic checkerboard CO state at the interface plane with GAF ordering for Mott-insulating side (CO-GAF). The phase boundaries, represented by the solid (broken) lines, correspond to the first-(second-) order phase transitions. In order to examine the role of the counter-ion layer thickness $N$, the thicker heterostructure, involving $N=10$ positive layers, is also investigated. The corresponding phase diagram is, however, analogous to the $N=4$ case apart from the AAF state, which is absent in the case of $N=10$. This type of $N$-dependence is further discussed in Sec.~\ref{subsec:canted AF}.

Intriguingly, the charge (magnetic) properties in the mean-field phase diagram are dramatically changed around $E_c/t \simeq 1$ ($U/t \simeq 3$). Let us first focus on the small $E_c$ regime (weak confinement regime $\lambda_{TF}>a$). At small $U$ the possible ground state is PMM. With increasing $U$, this state undergoes a continuous phase transition to the GAF state, whose behavior is in good agreement with Ref.~\onlinecite {okamoto:2005}; in spite of the lower occupancy $n\simeq 0.4$, the finite AF moment,  aligned in the same direction as that of the neighboring layers, is observed at the interface layer (see also Fig.~\ref{fig:ca_u12_dens}). Note that this GAF state is stabilized in the whole $E_c$ region at intermediate $U$. When $U$ is further increased, we encounter a distinct magnetic phase: a ferromagnetic component of the magnetization spontaneously arises around the hetero-interface, giving rise to a tilt of the AF moment. We, thus, arrive at the CA-GAF phase, although AAF states are still stable under certain conditions. The detailed study for the CA-GAF state is given in Sec.~\ref{subsec:canted AF}.

On the other hand, the charge sector exhibits quite different properties for $E_c/t >1$ ($\lambda_{TF}<a$). In this case, the PMCO state is stabilized in the small-$U$ region, which turns into the GAF state around $U/t\simeq 3$ via a first-order transition. Here, the charge-ordering is realized in the whole Mott-insulating region. We naturally expect that the PMCO state is stabilized by the long-range Coulomb interaction~\cite{PhysRevB.12.5249}. 
For the large $U$ region, we propose another interesting mechanism to stabilize the charge ordering in the CO-GAF phase, which will be addressed later in Sec.~\ref{subsec:charge order}.

In the following subsections, we give a more detailed discussions on the phase diagram with particular emphasis on the two remarkable phases realized in the large $U$ region: the canted antiferromagnetic state and the charge ordered state.


\subsection{Canted Antiferromagnetic state for $E_c/t <1$} \label{subsec:canted AF}

We start with discussions on the possible ordered states for $E_c/t <1$ ($\lambda_{TF}>a$). In this case, the weak attractive potential makes the conduction electron deeply penetrate into the band-insulating region, and, thus, the charge distribution near the interface is relatively sensitive to the values of $E_c$. Hence, the electronic phases for $E_c/t <1$ would be connected with the spatial variation of the charge distribution through the band-filling dependence of the typical bulk phases~\cite{hirsh:1985,potthoff:1995}. In this regime, we find three phases which are common to $N=4$ and $N=10$: PMM, GAF and CA-GAF states. Among them, the most salient phase we found is the CA-GAF state, because there is no analog in the bulk phase diagram~\cite{hirsh:1985,0295-5075-20-1-010}. This state is defined by the $z$-axis modulation of the antiferromagnetic spin alignment; around the interface, the alternating up and down spins are both tilted in the same direction, and, thus, a ferromagnetic moment perpendicular to the AF moment emerges. To obtain the above solution we consider the two kinds of the phases accompanied with finite magnetization around the interface, which are distinguished by the angle between the direction of the ferromagnetic and antiferromagnetic moment: perpendicular or parallel. With the energetic comparison, we find the solution with perpendicular one is favorable, and therefore the interplay of the F and AF orderings induces the spin canting in the vicinity of the interface.

\begin{figure}
\centering
\includegraphics[width=0.9\linewidth]{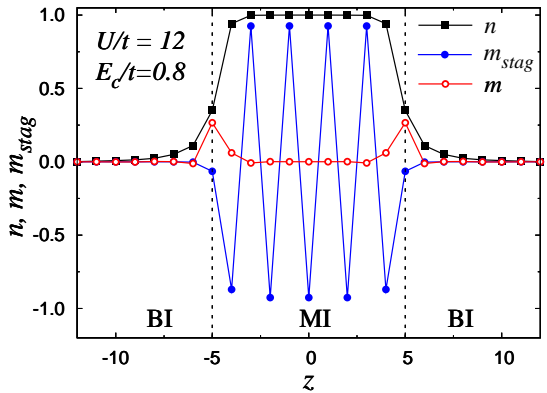}
\caption{Layer-dependent electron density (filled square), magnetization (open circle) and staggered magnetization (filled circle) for the parameters $U/t=12$, $E_c/t=0.8$, and $N=10$. Counter-ions are placed at $z=\pm 0.5, \pm1.5,\dots \pm4.5$. The interface layers are located at $z=\pm5$, separating Mott- and band- insulator regions.} 
\label{fig:ca_u12_dens}
\end{figure}

 In Fig.~\ref{fig:ca_u12_dens}, several quantities in the CA-GAF state are plotted as a function of the distance $z$ from the center of the heterostructure: the spatial modulation of the layer-dependent electron density $n(z)$, magnetization $m(z)$ and staggered magnetization $m_{stag}(z)$. This figure confirms the above statement that the weak charge binding $E_c<t$ ($\lambda_{TF}>a$) induces the charge leakage toward the band-insulating region; the electron density drops from $n\simeq 1$ to $\simeq 0$ over a few layers across the interface, and the finite electron density is even observed in outer layers ($ |z| > 5 $). In addition, the band occupation is always less than $n=1$ through the heterostructure. These features are analogous to the preceding studies~\cite{okamoto:2004:dmft, ruegg:2007, PhysRevB.74.195427, PhysRevB.74.075106}. Turning to the properties of magnetism, the values of the staggered magnetization are almost saturated inside the MI region ($ |z| < 5 $) and at the interface layers, respectively, and the magnetic moments are gradually diminished away from the corresponding regions. We note that in this CA-GAF phase, the characteristic behaviors of $n(z)$, $m_{stag}(z)$ and $m(z)$ presented in Fig.~\ref{fig:ca_u12_dens} are consistent with the DMFT results~\cite{okamoto:2005}, where good agreement between DMFT and Hartree -Fock analysis is only found for the electron density and the staggered magnetization. 

%
\begin{figure}
\centering
\includegraphics[width=0.9\linewidth]{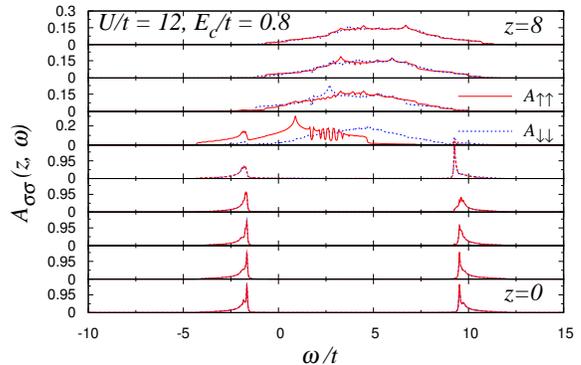}
\caption{Layer-resolved single-particle spectral function $A_{\sigma \sigma}$ for the CA-GAF state with $U/t=12, E_c/t=0.8, N=10$. Solid and broken lines indicate $A_{\uparrow \uparrow }$ and $A_{\downarrow \downarrow }$ components, respectively.}
\label{fig:ca_u12_dos}
\end{figure}
%

All such features manifest themselves in the corresponding layer-resolved single-particle spectral function $A_{\sigma, \sigma '}(z, \omega)$ in Fig.~\ref{fig:ca_u12_dos}, where we plot one of the spin-dependent components $A_{\sigma, \sigma}(z, \omega)$. Around the center of the heterostructure ($z\simeq0$), the conduction band gets quite narrow, and the opening of the gap ($\simeq Um_{stag}(z) \simeq U$) is observed due to the AF ordering; inside the MI region, the competition of the on-site $U$ and the attractive Coulomb potential sustains the electron occupancy approximately $1$ per site. In contrast, in the BI regions, the form of spectral function well reproduces that of the free electrons in the bulk systems \cite{okamoto:2004:dmft,okamoto:2005,ruegg:2007}, and the whole conduction band is finally pushed  above the Fermi level for $|z|\agt 7$. However, when approaching the interface, the spectral function is gradually shifted downward to the chemical potential because of the interplay of the electron spreading and binding effects along the $z$-axis. Metallic layers are, thus, formed around the interface, and as can be seen in Fig.~\ref{fig:ca_u12_dens}, these layers also carry a ferromagnetic spin polarization. Note that although the layer-dependent magnetization looks in good agreement with the previous DMFT analysis~\cite {okamoto:2005}, the origin of the present magnetization should be understood by the so-called Stoner mechanism. Actually the mean-field phase diagram of two-dimensional Hubbard model well describes the emergence of the spontaneous magnetization for $U/t\simeq 10$ at the interface showing $\simeq 0.4$ band occupation~\cite{hirsh:1985}. The small, but finite ferromagnetic moments in the neighboring layers could, thus, be identified as a proximity effect from the F ordering in the interface. Therefore, it is necessary to employ other sophisticated methods to give more accurate descriptions of the magnetism in the correlated heterostructure. These treatments are, however, beyond the scope of this paper, and we leave this issue for future research.

%
\begin{figure}
\centering
\includegraphics[width=0.9\linewidth]{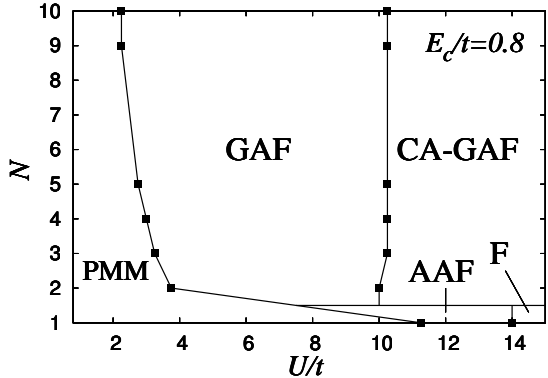}
\caption{Ground state phase diagram for the local Coulomb interaction $U$ and thickness of counter-ion layers $N$ at $E_c/t=0.8$. All transitions for each $N$ are found to be second order except for the first-order AAF-F phase boundary at $N=1$.}
\label{fig:u12ec08UNphase}
\end{figure}
%
We now return to the discussion on the CA-GAF state. As can be seen in Figs.~\ref{fig:ca_u12_dens} and \ref{fig:ca_u12_dos}, the gradual suppression of the spin canting is observed toward the center of the heterostructure, and accordingly the canted AF state is restricted to only a narrow region around the interface. Therefore, these results suggest the weak thickness-dependence for the stabilization of the CA-GAF state. This finding is supported by the obtained phase diagram with on-site $U$ and cation layer thickness $N$, which is presented in Fig.~\ref{fig:u12ec08UNphase}. For $N>2$, the general properties are similar; there are PMM, GAF and CA-GAF states from small to large $U$ regime. In contrast, the result of $N=1$ shows a considerably different behavior: the PMM state for small $U$, the AAF state for narrow range of intermediate $U$ and the F state for large $U$. Here, in the CA-GAF state at $N=2, 3$, all layers in the MI show the metallic density of states and a canted spin state, while such layers are limited around the interface for thicker heterostructures $N\agt4$. Compared with the PMM-to-GAF transition line, the phase boundary between the GAF and the CA-GAF state are less sensitive to $N$. As mentioned above, this tendency may be explained by the Stoner concept for metallic ferromagnets, because the electron occupancy at the interface layer is essentially independent of the layer thickness for large $U$~\cite{hirsh:1985}.

In contrast, if $E_c/t$ becomes smaller 
it is expected that the corresponding phase diagram shows a behavior more sensitive to the thickness $N$. This point is understood as follows; the weak charge binding reduces the occupancy at the center of heterostructure, and accordingly smaller $N$ further assists the electron leakage toward the band-insulating region and shows smaller occupation even at the central layers, as compared to thicker heterostructures. The calculated $U$-$E_c$ phase diagram in Fig.\ref{fig:n4_UEc_Phase} confirms the above statement. Actually, at $E_c/t \simeq 0.4$, we can find the AAF phase at $N=4$, which is not found at $N=10$. Unfortunately, however, it seems difficult for the present mean-field analysis to catch the physical aspect for such small $E_c$ region because the Stoner mean field treatment likely overestimates the ferromagnetic tendency in those layers~\cite{PhysRevB.56.R8479,0953-8984-3-26-014, okamoto:2005}. Hence, the study for such $E_c/t \ll 1$ regime will be given elsewhere.

\subsection{ Charge ordered state for $E_c/t>1$} \label{subsec:charge order}

%
\begin{figure}
\centering
\includegraphics[width=0.9\linewidth]{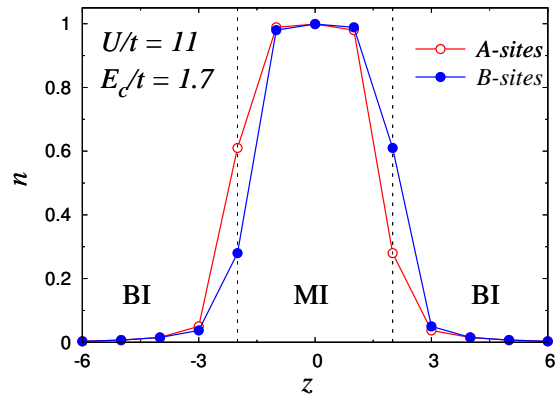}
\caption{Comparison of the sublattice-dependent electron densities for $U/t=11, E_c/t=1.7$ at a $N=4$ heterojunction as a function of $z$. Counter-ion planes are placed at $z=\pm 0.5, \pm 1.5$. Open and filled circles represent the electron occupations at A and B sublattices, respectively. 
}
\label{fig:co_u11ec17_dens}
\end{figure}
%
In this subsection, we focus on the regime $E_c/t>1$ ($\lambda _{TF}<a$) where electrons are strongly confined along the $z$ direction. Compared to $E_c/t<1$, the electronic phase diagram for $E_c/t>1$ is less sensitive to the layer thickness $N$ and the profile of the electron density. In the corresponding parameter regime, the strong attractive potential traps the electrons inside the MI region, and accordingly the most mobile electrons are absent in the band-insulating region. This observation suggests that the role of the interlayer hopping, i.\ e.\ electron hopping along the $z$ direction, is distinguished from that of in-planes, particularly around the interface. It is because the layer-dependent electron occupancy shows an abrupt change across the interface: from $n \simeq 1$ in the Mott-insulating region to $n\simeq 0$ in the band-insulating region. In the region of $E_c/t>1$, the corresponding phase diagram in Fig.~\ref{fig:n4_UEc_Phase} typically consists of the COPM, GAF and CO-GAF state. Note that the stabilization of the COPM phase as well as GAF state is well understood via the bulk phase diagram~\cite{PhysRevB.12.5249}. So, we hereafter focus on the interface-specific CO-GAF state. 

\begin{figure}
\centering
\includegraphics[width=0.9\linewidth]{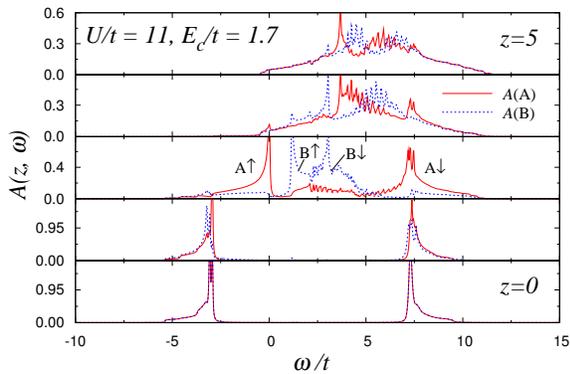}
\caption{Sublattice-dependent layer-resolved spectral function of CO-GAF state at $U/t=11, E_c/t=1.7$ for the $N=4$ heterostructure. 
Solid and dashed line represent the spectral functions for A- and B-sublattices, respectively, and $\sigma=\uparrow,\downarrow$ illustrate the corresponding spin state for each sublattice. }
\label{fig:co_u12ec17_dos}
\end{figure}

The sublattice-dependent charge distribution for the CO-GAF state is shown in Fig.~\ref{fig:co_u11ec17_dens}. Most of the electrons reside inside the heterostructure, and, thus, only a few electrons are found in the outer layers. In addition, the resulting checkerboard-type charge ordering can be found only at the interface layer, which is accompanied by a ferromagnetic spin configuration. Therefore, the present result is consistent with the previous density functional calculation performed by Pentcheva and Pickett~\cite{pentcheva:2007}. The corresponding layer-resolved spectral function $A(\omega)$ is presented in Fig.~\ref{fig:co_u12ec17_dos}, where $A($A$)$ (solid lines) and $A($B$)$ (dotted lines) indicate the resulting spectral functions for the A and B sublattices, respectively. Due to the interface charge-ordering, a gap is formed at the interface layer ($z=2$), which would suppress the metallic interface state with approaching $E_c/t \simeq 2.4$, where the insulating interface is obtained as in Ref.~\onlinecite{pentcheva:2007}.

%
%

\begin{figure}
\centering
\includegraphics[width=0.75\linewidth]{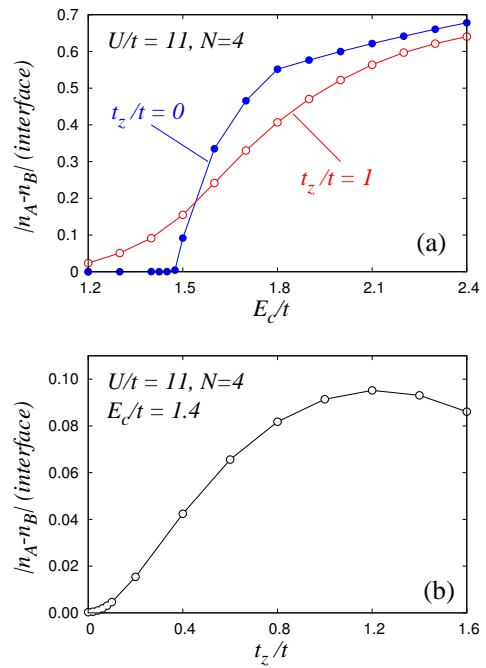}
\caption{Order parameter $|n_A-n_B|$ of the interface charge-ordering for the $N=4$ heterojunction with interface electron occupancy at A(B)-sublattice $n_A$($n_B$). (a) $|n_A-n_B|$ is illustrated as a function of $E_c$ for $U/t=11$. Open and filled circle denote the resulting isotropic ($t_z/t=1$) and anisotropic ($t_z/t=0$) electron transfer, respectively. (b) the detailed $t_z$-dependence is depicted for $E_c/t=1.5$. To obtain the values for $t_z/t \not= 1$ we here use the result of $t_z/t=1$ as an initial input and tune the magnitude of the inter-layer hopping.}
\label{fig:co_u11n4_tz}
\end{figure}

We now examine possible driving forces for the interface charge-ordering. Figure~\ref{fig:co_u11n4_tz}(a) shows the order parameter $|n_A-n_B|$ of the ferromagnetic CO phase  as a function of $E_c$. Here, $n_A$ ($n_B$) represents the interface electron density at the A (B) sublattice. The filled and open circles show the results for different values of interlayer hopping $t_z$.  At $t_z/t=0$, the electron motion is restricted in the two-dimensional sheet while $t_z/t=1$ corresponds to the isotropic one. 
We find that the interface charge-ordering is induced in both cases. Note that the charge-ordering is stabilized even at $t_z/t =0$ in the large $E_c$ region, implying that the ferromagnetic CO state is stabilized due to the long-range Coulomb repulsion within the same plane.~\cite{pentcheva:2007}
Namely, the interface layer has the electron density $n \simeq 0.4$, which increases with increasing $E_c$ and finally approaches the quarter filling where the checkerboard charge-ordering is particularly favorable. 

%
%

On the other hand, we should take into account the effects of electron transfer in the $z$-direction to explain the charge order emerging for $t_z/t=1$ in the region of $E_c/t<1.5$ in Fig.~\ref{fig:co_u11n4_tz}(a). Figure~\ref{fig:co_u11n4_tz}(b) presents the detailed study of the $t_z$-dependent CO parameter at $E_c/t=1.4$. As seen from the figure, the CO phase becomes more pronounced with increasing $t_z$. This trend seems, however, to contradict the above-mentioned mechanism, since the increase in $t_z$ causes the electron transfer into the band-insulating region and, thus, depletes the interface, potentially suppressing charge ordering. Therefore, it seems difficult to attribute the origin of the CO-GAF state only to the long-range Coulomb repulsion for $E_c/t \simeq 1.5$. 

%
\begin{figure}
\centering
\includegraphics[width=0.8\linewidth]{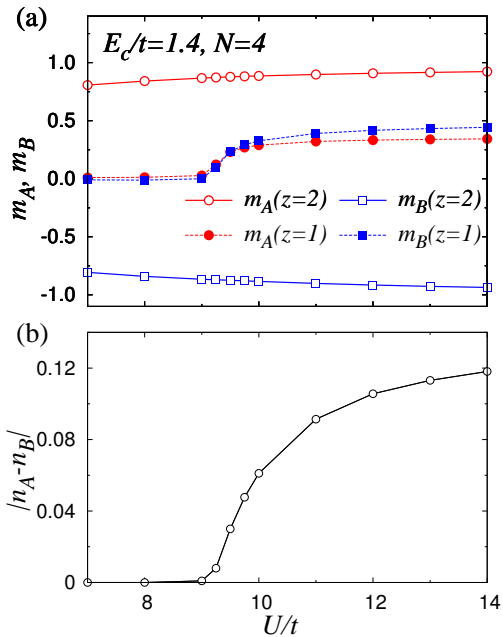}
\caption{Plot of $U$-dependence of several quantities near the interface around the transition point from GAF to CO-GAF state. (a) the magnetization for A-sublattice (squares) and B-sublattice (circles) at $E_c/t=1.4$. Open and filled symbols refer to the values of the interface ($z=2$) and the neighbor layer ($z=1$), respectively. (b) corresponding order parameter of CO state at interface layer.}
\label{fig:co_ec14n4}
\end{figure}
%

To elucidate another possible mechanism, we focus on a nontrivial role of $t_z$ that induces the effective interlayer magnetic interaction.  Figures \ref{fig:co_ec14n4}(a) and (b) show the sublattice-dependent magnetization and the interface CO parameter as a function of $U$, respectively. In the vicinity of the CO-transition point $U/t \simeq 9.5$, the interface magnetization spontaneously appears via the exchange mechanism, causing different spin configurations between two sublattices; A-sites (lower occupancy) present a ferromagnetic spin alignment with the neighbor layer located inside the heterostructure while B-sites (higher occupancy) prefer the anti-parallel one. Hence, ferrimagnetism is stabilized at the interface. 
According to our numerical results, A-sites always carry the lower electron density compared to B-sites around $z\simeq 2$. To clarify the relation between this spin structure and the interface charge-ordering, we take into account the charge confinement effect along the $z$-axis; for $E_c/t>1$ (i.e. $\lambda_{TF}<a$), the strong charge binding via the positive background localizes the electrons inside the heterostructure ($z\leq |3|$ in Fig.~\ref{fig:co_u11ec17_dens}), and thus it is not energetically favorable to keep the higher occupancy outside the heterostructure. Therefore,  the electron occupancy for the A-sites decreases to suppress the charge-transfer toward the band-insulating region. 
 This in turn increases the occupancy at B-sites, which may enhance the antiferromagnetic correlation along the $z$ direction. This mechanism based on the interlayer spin couplings naturally explains the behavior of the CO parameter for small $t_z$ in Fig.~\ref{fig:co_u11n4_tz} (b).  
In summary, we may interpret the interface charge order as being induced by interlayer spin coupling in order to reduce the charge leakage from the interior toward the band insulator. This spin-driven mechanism, together with long-range Coulomb interaction, stabilizes the CO-GAF phase even for the relatively weak $E_c$.

\section{Effects of magnetic fields} \label{effects of magnetic fields}

In this section, we discuss the nature of the heterostructure under external magnetic fields to further clarify the interplay of the magnetism and the charge density profile. As discussed in the previous section, the possible ground states are closely connected with the profile of charge distribution.
 At zero magnetic field, the long-range Coulomb interaction $E_c$ plays an important role in determining the charge density profile along the $z$-axis. We also note that magnetic fields can have a strong influence on the charge distribution by modifying in the spin structure which governs the charge distribution in the vicinity of the interface~\cite{PhysRevB.74.075106}. In the following, we analyze the effect of magnetic fields on  the density modulation. Throughout this section, we consider the heterostructure with $N=10$, but the obtained results apply to a wider range of $ N $ (see Sec.~\ref{subsec:meta}). We focus on the relatively large $U$ region in which the GAF state is stabilized inside the heterostructure, and consider the magnetic fields perpendicular to the AF-moments using the Zeeman coupling term. Under these circumstances, we will deal with three phases, i.\ e.\ GAF, CA-GAF and CO-GAF.

\subsection{Metamagnetic transition} \label{subsec:meta}

\begin{figure}
\centering
\includegraphics[width=0.9\linewidth]{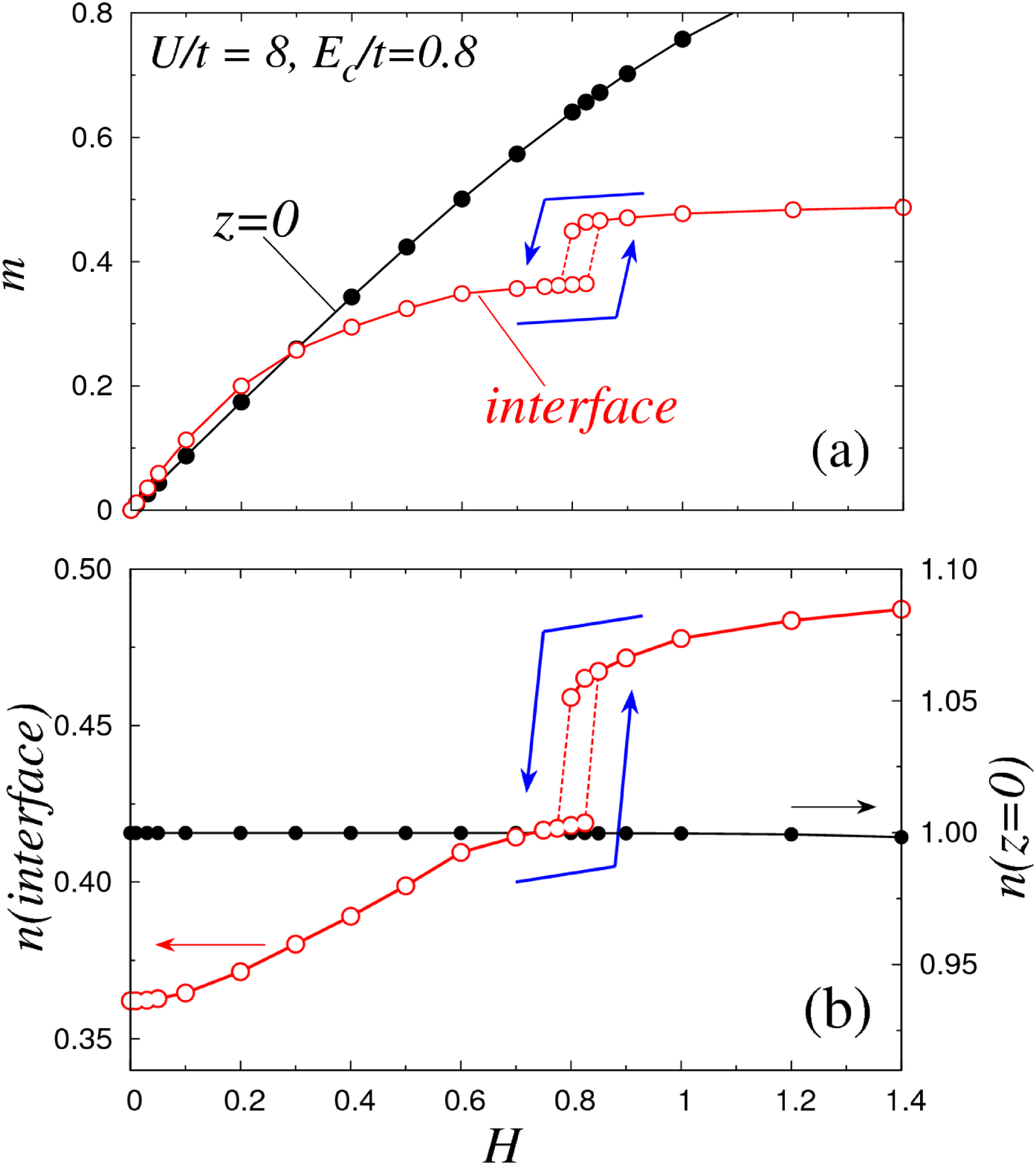}
\caption{ (a) Comparison of magnetization curves at the interface (open) and at the center of the heterostructure $z=0$ (filled symbol) with the fixed values $U/t=8$ and $E_c/t=0.8$. (b) The corresponding electron density at the interface (left axis) and $z=0$ (right axis). The up (down) arrow, denoting the hysteresis loop, presents the magnetization under the increase (decrease) of magnetic fields. All quantities are computed under the magnetic fields perpendicular to the AF moment in the Mott-insulating region.}
\label{fig:meta_u8ec08_h}
\end{figure}

We first concentrate on the appearance of a meta\-magnetic transition in the heterostructure under external magnetic fields. Figure~\ref{fig:meta_u8ec08_h}(a) shows the layer-dependent magnetization as a function of $H$ for fixed values $U/t=8$ and $E_c/t=0.8$. Note that in the absence of external fields, the corresponding ground state is the GAF in this parameter regime (Sec.~\ref{sec:phase diagram}).  With increasing $H$ the interface magnetization (open circles) displays a discontinuous jump around $H\simeq 0.8$, signaling a first-order metamagnetic transition. This is confirmed by a hysteresis loop emerging with rising and lowering magnetic fields. In contrast, we do not find any anomalous behaviors in the magnetization at the central layer (filled circles); it increases monotonically without exhibiting any jumps, and smoothly approaches the saturated value $m\simeq n \simeq 1$. These results imply that the metamagnetism observed here is intrinsic to the interface. Interestingly, this metamagnetic transition is intimately connected with the charge distribution, as shown in Fig.~\ref{fig:meta_u8ec08_h}(b), where the left (open) and right (filled symbols) axes display the computed electron density at the interface and the central layer, respectively. As clearly seen in this figure,  the electron density exhibits the same anomalous behavior as in the magnetization in Fig.~\ref{fig:meta_u8ec08_h}(a): (i) there exist the abrupt jumps in both quantities only at the interface at the same magnetic fields and (ii) a hysteresis loop emerges with increasing/decreasing magnetic fields. This pronounced change in the electron density is restricted to the vicinity of the interface.

%
%

\begin{figure}
\centering
\includegraphics[width=0.8\linewidth]{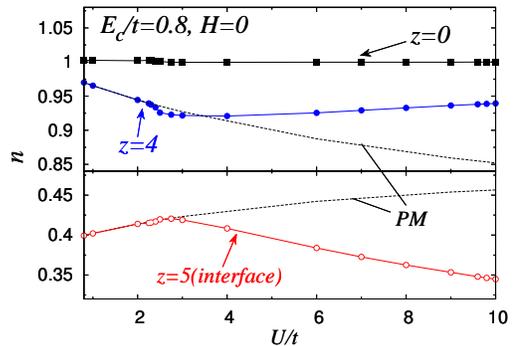}
\caption{The $U$-dependent electron density for given layers labeled by $z = 0, 4, 5$ at $E_c/t=0.8$ and zero magnetic field (see Fig.~\ref{fig:ca_u12_dens}): broken lines for the paramagnetic state at the interface ($z=5$, lower panel) and for the neighboring plane to the interface ($z=4$, upper panel).
 For layer indices, see Fig.\ref{fig:ca_u12_dens}.
}
\label{fig:meta_ec08_dens}
\end{figure}

Therefore, in order to understand the origin of the interface metamagnetism, it is crucial to elucidate how the magnetism is related to the charge distribution in the heterostructure. To this end, we show the $U$-dependent electron density for $H=0$ at several layers  in Fig.~\ref{fig:meta_ec08_dens}: the center layer ($z=0$), the nearest-neighbor layer to the interface  ($z=4$) and the interface layer ($z=5$). The characteristic feature appears at the PMM-to-GAF transition point $U/t \simeq 3$, around which the gradients of the curves for $z=4$ and $z=5$ change their signs, in contrast to the monotonic change in the PMM state (dotted lines). On the other hand,  the electron density at the central layer depends only weakly on $U$. This observation is consistent with the previous finding that the density profile is affected by the electronic phases inside the heterostructure~\cite{ruegg:2007, PhysRevB.74.075106}. Particularly in Ref.~\onlinecite {ruegg:2007}, R\"uegg {\it et al.} suggested that the visible modification occurs via the localization of the electronic states with undergoing a Mott transition; mobile electrons are tightly bounded inside the heterostructure to keep the higher band-occupation  $\simeq 1$. This scenario is also applicable to the present case since in the bulk phase diagram, the conventional GAF phase prefers the singly occupied sites. Therefore, in order to support the emergence of antiferromagnetism, it is necessary to confine the electrons inside the heterostructure with varying the density profile along the $z$-axis. We note that such charge redistribution occurs only near the interface~\cite{PhysRevB.74.075106}, and plays a rather minor role in comparison with the long-range interaction $E_c$. Hence, the interface electrons are mainly confined to the Mott-insulating region, and the resulting density profile changes abruptly across the interface.
%
%

Recall here that the applied magnetic fields have a tendency to destroy the AF ordering. Such a suppression seems to be stronger near the interface than the central layers because the number of the nearest-neighbor sites coupled antiferromagnetically to each other decreases from $6$ at the central layers to $5$ at the interface. This, in turn, increases the field-induced magnetic moment around the interface. Especially the interface magnetization is immediately saturated because of the extremely weak AF ordering and the large spin susceptibility via in-plane instability toward Stoner ferromagnetism. Nevertheless, below the metamagnetic transition point, the interface AF-moment remains finite through the proximity effect from the neighboring layers (see Fig. \ref{fig:ca_u12_dens}). Thus, it will be energetically favorable in a magnetic field to increase the electron occupation at the interface. On the other hand, the charge confinement along the $z$-axis, driven by magnetism, would become more relevant for large $U$ because of the strong AF coupling as shown in Fig.~\ref{fig:meta_ec08_dens}. 
This means that the competition between correlation effects and the influence of the magnetic field is crucial for determining the AF order at the interface.  This gives rise to the metamagnetic transition accompanied by a redistribution of electron charge near the interface. At the critical magnetic field, we indeed find the discontinuous drop in the staggered magnetization near the interface, confirming that the metamagnetism is indeed caused by the charge redistribution of electrons, as seen in Fig.~\ref{fig:meta_ec08_dens}.

\begin{figure}
\centering
\includegraphics[width=0.8\linewidth]{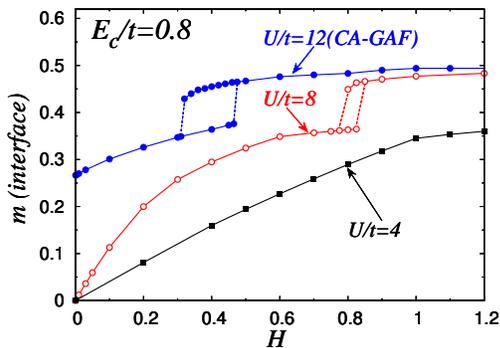}
\caption{The magnetization curves at the interface for several values of the on-site interaction. The corresponding ground states at $H=0$  are GAF for $U/t = 4$ (filled squares), $8$ (open circles), and CA-GAF at $U/t=12$ (filled circles), respectively.}
\label{fig:meta_ec08_uh}
\end{figure}

This view is further confirmed by our numerical results. Figure~\ref{fig:meta_ec08_uh} shows the computed magnetization at the interface for several values of $U$. The resulting magnetization remarkably changes its behavior across a critical end point of $U/t\simeq 7.5$. Though the magnetization varies continuously below this value, the first-order metamagnetic transition is obtained for $U/t \agt 8$, above which the critical magnetic field displays a gradual decrease with increasing $U$. These tendencies agree well with the above suggestion, because relatively large values of $U$ may enhance the difference between two effects: spreading and confinement of electrons along the $z$ direction due to magnetic fields and local Coulomb interaction, respectively.
%
%

\subsection{Field induced charge ordered state} \label{subsec:field_co}

\begin{figure}
\centering
\includegraphics[width=0.8\linewidth]{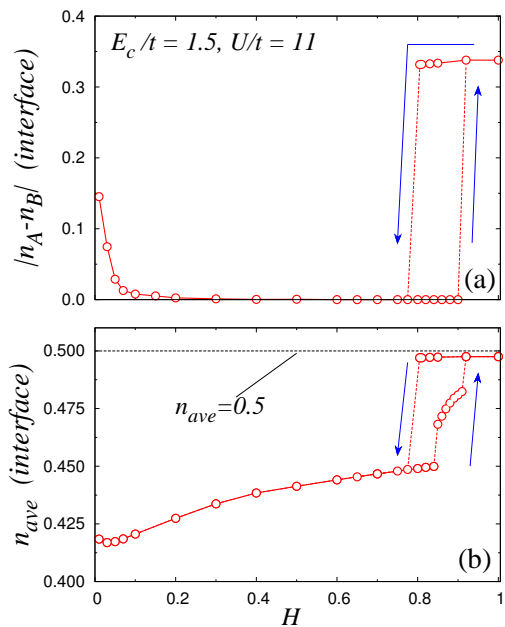}
\caption{Plot of interface electron density at $U/t=11, E_c/=1.5$ as a function of magnetic fields. (a) order parameter of the checkerboard charge-ordering.  The increase (up arrow) and decrease (down arrow) of the magnetic fields yield a hysteresis loop. (b) the corresponding interface occupation number, defined by $n_{ave}=(n_A+n_B)/2$. Dotted line shows the value of the quarter filled band: $n_{ave}=1/4$.}
\label{fig:field_co_u11ec15_h}
\end{figure}

In the relatively large $E_c$ regime, 
the charge redistribution under external magnetic fields reveals another interesting aspect of the correlated heterostructure: field-induced charge ordering. Generally, the strong inter-site Coulomb interaction ($\simeq E_c$) may break the translation invariance and induce the CO in the bulk systems. As mentioned in the previous section, at $H=0$, there actually exists the interface CO state of checkerboard pattern, i.e. CO-GAF phase.\cite{pentcheva:2007} Figure~\ref{fig:field_co_u11ec15_h}(a) displays the behavior of such interface charge ordering in applied magnetic fields. In weak magnetic fields, the CO immediately disappears with increasing field. This is caused by the emergence of the special magnetic order at the interface which partially supports the stabilization of the interface charge-ordering as mentioned in Sec.~\ref{subsec:charge order}, so that the disappearance of the magnetic order weakens the interface charge-ordering. 

Surprisingly, we encounter a reentrant CO state with further increasing $H$; the checkerboard charge-ordering emerges again around $H \simeq 0.9$ only at the interface. Note that there is a discontinuous jump accompanied by a hysteresis loop in the order parameter, implying that the transition is of first order. Furthermore the order parameter of the field-induced CO phase is almost unchanged once it is induced. This $H$-dependence suggests that the origin of the interface charge-ordering is the inter-site Coulomb repulsion. This view is indeed confirmed in  Fig.~\ref{fig:field_co_u11ec15_h}(b). Above $H\simeq 0.9$, around which the field-induced CO phase occurs, the averaged electron density at the interface $n_{ave} =(n_A +n_B)/2 $ shows a plateau at $n_{ave}=0.5$. As is the case for a conventional GAF state, the checkerboard CO can be caused by the Fermi-surface nesting, particularly at quarter filling. Therefore, the in-plane CO state stabilized with $n_{ave}=0.5$ above the critical magnetic field in Fig.~\ref{fig:field_co_u11ec15_h} evidences that this CO state is driven by the intersite Coulomb repulsion under the quarter-filling condition. We note in Fig.~\ref{fig:field_co_u11ec15_h}(b) that below the critical magnetic field ($H\sim 0.9$), there is another first-order transition around $H\sim 0.85$ with a small jump in $n_{ave}$. This is due to the metamagnetic transition discussed in Sec.~\ref{subsec:meta} (see also Fig.~\ref{fig:N10_UEc_Phase}(a)). 

%
%



Since there exists a finite charge excitation gap at the interface, the field-induced CO state would be experimentally detected, for example, as a rapid drop in the conductivity with increasing $H$. 

\subsection{Phase diagram under magnetic fields} \label{subsec:h_phase}
%
\begin{figure}
\centering
\includegraphics[width=0.85\linewidth]{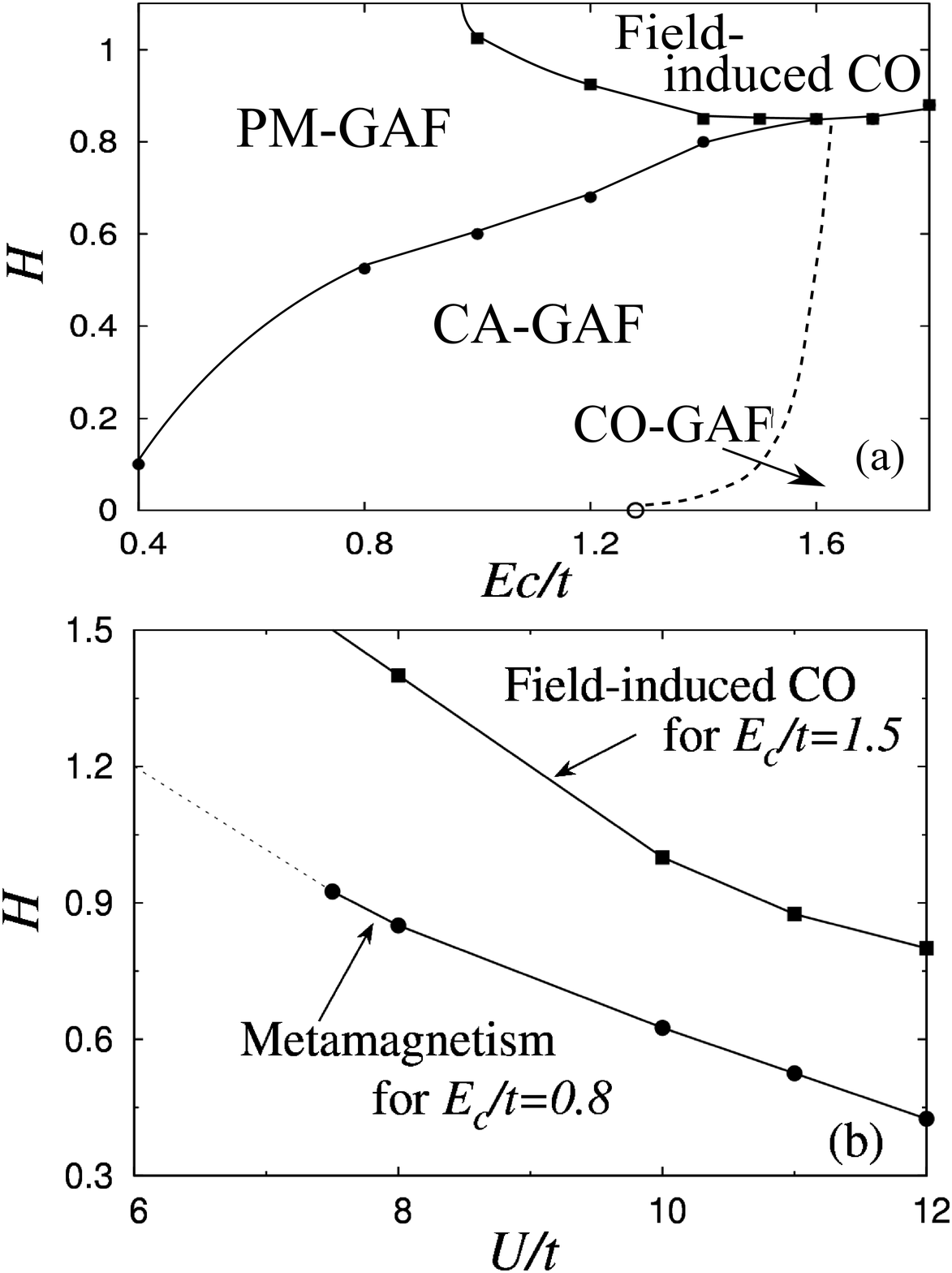}
\caption{The obtained phase diagrams under external magnetic fields: (a) $H$-$E_c$ phase diagram for $U/t=11$. Open circles: CO- and CA-GAF phase boundary at zero-magnetic field; filled circles: the first-order metamagnetic transition from CA-GAF to GAF state with paramagnetic interface (PM-GAF),; filled squares: first-order phase boundary of the field-induced CO state. As well known, the PM-GAF state turns into the paramagnetic phase for large $H$.
(b) $H$-$U$ phase diagram with different values of $E_c$. Filled circles: the first-order metamagnetic transition line for $E_c/t=0.8$, across which GAF (or CA-GAF) turns into PM-GAF with the increase in $H$; the characteristic feature is, however, changed to continuous one for $U/t\agt 7.5$. Filled squares: the phase boundary obtained at $E_c/t=1.5$, illustrating the discontinuous transition toward the field-induced CO phase with increasing $H$.}
\label{fig:N10_UEc_Phase}
\end{figure}
%

It is instructive now to present the phase diagram under magnetic fields. Figure~\ref{fig:N10_UEc_Phase}(a) shows the calculated phase diagram with the parameter $E_c$ and the magnetic field $H$. The field-induced charge-ordering (closed squares) can be found in the regime of relatively large $E_c$ and $H$. This trend indeed confirms our prediction presented in Sec.~\ref{subsec:field_co} concerning the origin of the transition: the important interplay of the  band occupancy ($n_{ave}=0.5$) and the long-range Coulomb repulsion. In contrast, increasing $E_c$ seems less relevant for the metamagnetic transition (filled circles), since this transition is accompanied with the increase of electron density at the interface,
which is unfavorable for large $E_c$. 
Interestingly, such charge redistribution would further enhance the tendency toward the field-induced CO transition. Indeed these transition lines gradually merge in the large $E_c$ region.

\begin{figure}
\centering
\includegraphics[width=0.9\linewidth]{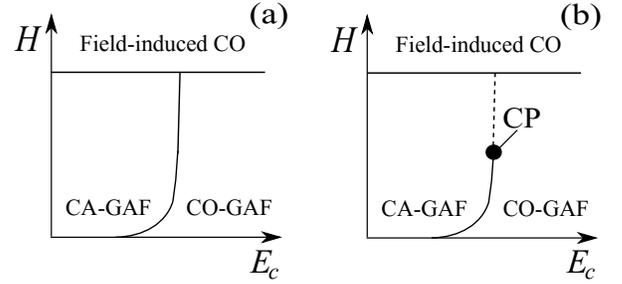}
\caption{Schematic $H$-$E_c$ phase diagram with particularly focusing on the boundary between CO-GAF and CA-GAF phases. Solid lines: first-order transition, broken line: second-order transition. Thus, (a) CO- and CA-GAF states are separated by a discontinuous transition, while (b) there exists a critical point (CP), above which these phases are connected continuously.}
\label{fig:field_co_transition}
\end{figure}

The nature of phase transition from CA-GAF to CO-GAF state under the magnetic fields, shown by the dotted line in  Fig.~\ref{fig:N10_UEc_Phase}(a), remains still open. At intermediate $E_c$ ($\simeq 1.4t$), the specific interlayer magnetic coupling might drive the CO-GAF phase at $H=0$ [Sec.~\ref{subsec:charge order}]. Therefore, if such a spin structure is suppressed with the increase in $H$, the CO-GAF state would undergo a first-order transition to the CA-GAF state as is the case for $H=0$. Note that these two phases are distinguished by the direction of the interface F-moment as well as the CO parameter, $|n_A-n_B|$. However, for large $E_c\agt 1.6t$, the interface charge-order even survives in the large $H$ region without the spin coupling between layers, and thus the difference between CO- and CA-GAF states are only given by the order parameter $|n_A-n_B|$. Accordingly, as shown in Fig.~\ref{fig:field_co_transition} (b), the CO- and CA-GAF phases might be continuously connected beyond the certain critical value of $H$. In the present study, however, we cannot determine which of the two scenarios, depicted in \ Fig.~\ref{fig:field_co_transition} (a) and (b), is correct. 
%
%

Figure~\ref{fig:N10_UEc_Phase}(b) presents the $U$-dependence of the meta\-magnetism and the field-induced charge-ordering, respectively. As mentioned in Sec.~\ref{subsec:meta}, the metamagnetic transition can appear only in the large $U$ region, and thus a critical point may exist around $U/t \simeq 7.5$. Here, the overall $U$-dependence of the field-induced charge-ordering resembles that of the metamagnetic transition; the value of the critical magnetic field $H_c$ increases monotonically with decreasing $U$. However, the difference between these transitions is found in the lowest critical values of $H_c$. Although the interface metamagnetism transition is of first-order only for $U/t\agt 8$, we find the discontinuous charge-ordering transition in a much wider range of the on-site interaction. This is because the intersite Coulomb interaction with the specific occupancy of electrons plays a dominant role in the transition to the field-induced CO phase, and the on-site Hubbard interaction merely modifies the electron occupancy at the interface with inducing magnetism.

\section{Summary and Conclusion} \label{sec:conclusion}

In this paper, we have considered the generalized Hubbard model and presented the mean-field study on the ground-state properties of the correlated heterostructure composed of a Mott insulator (MI) sandwiched between two band insulators (BI). The model structure is electrostatically characterized by the positively charged ions located in the Mott-insulating region, causing a spatial modulation of the electron distribution. Such a profile of electron density along $z$ direction can be tuned by the long-range Coulomb interaction and also by the one-site interaction via the magnetic moment induced inside the heterostructure.
Using a relatively wide range of parameters and tuning spin structures via external magnetic fields, we have investigated how the inhomogeneous charge distribution leads to the intriguing electronic and magnetic phases realized in the model heterostructure.

At zero magnetic field, our model exhibits rich electronic phases in the $U$-$E_c$ plane, including CA-GAF, CO-GAF and some other conventional ordered states. The resulting phase diagram changes its charge character around $E_c/t\simeq 1$ ($\lambda_{TF}\simeq a$). For $E_c/t < 1$, where the density profile of electrons gradually changes beyond the interface layer, we have found the CA-GAF phase for relatively large $U$, which consists of the GAF oder inside the MI region and a canted AF phase in the vicinity of the interface. Interestingly, even though the spatially modulated magnetization is caused by the Stoner mechanism, our results are in good agreement with the earlier DMFT study of Okamoto and Millis~\cite {okamoto:2005}. Note that the CA-GAF phase has been proposed for the first time in this paper .
On the other hand, for $E_c/t>1$, we have found the ferromagnetic CO state with checkerboard pattern at the interface as suggested by previous {\it ab initio} calculations\cite{pentcheva:2007}. In some parameter regime, however, this CO state might not be explained only in terms of the long-range Coulomb repulsion. As an additional mechanism supporting the interface charge-ordering, we have proposed the importance of the interlayer electron hopping, which causes the site-dependent interlayer spin couplings and thereby enhances the instability toward the charge-ordering together with the long-range Coulomb repulsion. 

It has been demonstrated that a strong coupling between spin and charge degrees of freedom is crucial for the system in magnetic fields, giving rise to the field-driven charge redistribution in the vicinity of the interfaces. One of the remarkable consequences of this effect is a first-order metamagnetic transition. It has been shown that this transition is induced by the competition of two effects: the localization of electrons inside the center region of the heterostructure via the GAF ordering ($U$), and the charge transfer toward the interface via the spin polarization ($H$). This mechanism is fully supported by the computed $U$-dependence of the metamagnetic transition and the specific jump in the interface electron density as a function of $H$. 

Another remarkable phenomenon we found here is the field-induced charge-ordering transition. This transition at the interface is possible only for relatively large $E_c$, and is explained in terms of the modification of interface electron density through the competition of $U$ and $H$; the intersite Coulomb repulsion together with an emergent quarter-filled band condition gives the instability toward the CO phase. This also demonstrates the importance of the interplay between the spin and charge degrees of freedom in magnetic fields.
In order to gain further insights into the effects of magnetic fields, we have presented the $E_c$-$H$ and $U$-$H$ phase diagrams, and found that characteristic features of the resulting phase boundaries are consistent with our scenario for the field-driven metamagnetic/charge-ordering transitions.

In this paper, we have performed the mean-field analysis for the magnetic properties of the heterostructure, which are strongly coupled to the inhomogeneous charge distribution. Although this simple treatment provides an insight into physical properties of the model, it is particularly important to confirm the present results by taking into account strong electron correlations with many-body techniques such as DMFT~\cite{okamoto:2004:dmft} and slave-boson method~\cite{ruegg:2007}. This issue will be studied elsewhere.


\section*{Acknowledgments}
We are grateful to M. Ossadnik, A. R\"uegg, S. Pilgram and D. M\"uller for many helpful discussions. 
This research is granted by the Japan Society for the Promotion of Science (JSPS) through the ``Funding Program for World-Leading Innovative R\&D on Science and Technology (FIRST Program)'', initiated by the Council for Science and Technology Policy (CSTP). We also acknowledge the support by the Grant-in-Aid for Scientific Research [Grant nos. 21540359, 20102008] and the Global COE Program ``The Next Generation of Physics, Spun from Universality and Emergence'' from MEXT of Japan. We are also grateful for financial support by the Swiss Nationalfonds and the NCCR MaNEP.


\end{document}